\documentclass[twocolumn,preprintnumbers,amsmath,amssymb]{revtex4}
\usepackage{graphicx}
\usepackage{dcolumn}
\usepackage{bm}
\usepackage{amsmath}
\usepackage{amssymb}
\usepackage{amsthm}
\usepackage{amsfonts}
\usepackage{enumerate}
\usepackage{latexsym}
\usepackage{hyperref}
\usepackage{tikz}
\begin{document}

\title{Noncommutative Chern-Simons theory and exotic geometry emerging from the lowest
Landau level}

\author{Xi Luo$^{1}$} \email{xiluo@itp.ac.cn},

\author{  Yong-Shi Wu$^{2,3,4}$, and Yue Yu$^{2,3,1}$}

\affiliation {${}^1$Key Laboratory of Theoretical Physics, Institute of
Theoretical Physics, Chinese Academy of Sciences, P.O. Box 2735,
Beijing 100190, China\\
${}^2$ Center for Field
Theory and Particle Physics and State Key Laboratory of Surface Physics, Department of Physics, Fudan University, Shanghai 200433,
China \\
${}^3$Collaborative Innovation Center of Advanced Microstructures, Fudan University, Shanghai 200433, China\\
${}^4$Department of Physics and Astronomy,
University of Utah, Salt Lake City, Utah 84112, USA
 }

\date{\today}

\begin{abstract}
We relate the collective dynamic internal geometric degrees of freedom to the gauge fluctuations in $\nu=1/m$(m odd) fractional quantum Hall effects. In this way, in the lowest Landau level, a highly nontrivial quantum geometry in two-dimensional guiding center space emerges from these internal geometric modes. {Using the Dirac bracket method, we find that this quantum geometric field theory is a topological non-commutative Chern-Simons theory.}
Topological indices, such as the guiding center angular momentum (also called the shift) and the guiding center spin, which characterize the fractional quantum Hall (FQH) states besides the filling factor, are naturally defined. A noncommutative K-matrix Chern-Simons theory is proposed as a generalization to a large class of Abelian FQH topological orders.
\end{abstract}


\maketitle

\section{Introduction}

The discovery of the fractional quantum Hall effects (FQHE) delivered a new area of condensed matter physics, the two-dimensional strongly-correlated electron physics \cite{tsg}. After Laughlin's trial wave function \cite{laughlin}, several phenomenological effective theories were proposed for the FQHE \cite{effth1, effth11, effth2}. In particular, Wen presented a { pure} Abelian Chern-Simons(CS) theory to describe the topological order in a number of FQH states \cite{effth2}. In view of the need of the lowest Landau level (LLL) projection, noncommutative Chern-Simons theory \cite{nccs} seems to be a candidate for better description. However, the microscopic origin of the Chern-Simons gauge fluctuations is not yet clear.  

In this paper, we will show that the Chern-Simons gauge fluctuations originate from the collective fluctuations in electron position around the guiding center (which will be defined in the next section). These collective fluctuations that give rise to a fluctuating geometry were recently noticed by Haldane \cite{haldane}. Since the guiding center coordinates become noncommuting in the lowest Landau level\cite{haldanewu}, we further {show} 
that a highly nontrivial two-dimensional quantum geometry emerges in the guiding center space from these collective dynamic internal geometric degrees of freedom, and is 
{consistent with the} noncommutative Chern-Simons gauge theory {to the first-order expansion of the noncommutative parameter}. The latter is shown to provide additional topological observables naturally in distinguishing different fractional quantum Hall states besides the filling factor \cite{haldane}.

The attempt of finding a geometric description for FQH liquid was inspired by the study of Hall viscosity\cite{ASZ,read}. 
Haldane noticed that the Galilean metric $g_G^{ab}$ of the electron band mass may be generically different from the Coulomb metric $g^{ab}_c$  of the unscreened Coulomb potential in cases lacking the rotational symmetry. Then the Laughlin state for the FQHE is determined by pseudo-potentials given in the background of a variational metric $g_0^{ab}$ that interpolates between $g_G^{ab}$ and $g_c^{ab}$, which makes the correlation energy minimal.  The metric fluctuation $\delta g^{ab}({\bf x},t)$ is identified as the collective dynamic internal geometric degrees of freedom \cite{haldane}. An important question in this approach is how to formulate the dynamics that governs the geometric fluctuations in the effective theory at long distances. The main difficulty is the fact that after the lowest Landau level projection, the guiding center coordinates become noncommutative, as emphasized in \cite{haldanewu} or in the book \cite{liwu}. In this paper we report our results on this topic, in an approach that deals with the noncommutativity of guiding center coordinates in a way different from Haldane's \cite{haldane}. We show that these collective dynamic internal geometric degrees of freedom give rise to a (two-dimensional) quantum geometry.

We study the FQHE in the lowest Landau level using the zero band mass limit, namely $m_b\to 0$. The advantage of taking this limit is that we do not need to face the complications due to the derivatives appearing in the guiding center operator. The $m_b\to 0$ limit imposes a second-class constraint that the kinematic momentum $ \pi_b =0$. Upon quantization, conventional commutators are replaced by the Dirac brackets, and the electron coordinates (in two dimensions) become non-commutative under the Dirac brackets \cite{jac}.  The kinetic energy term in the Hamiltonian is set to zero when $m_b\to 0$, but there are residual collective quantum fluctuations of the electron position which are described by the topological quantum mechanics with a pure Chern-Simons Lagrangian\cite{jac}. For a many-body system, in the continuum limit we write the position field of electrons to be the sum of the guiding center position field and the position fluctuation around the guiding center, the effective Lagrangian can be transformed into an Abelian Chern-Simons theory in the two-dimensional guiding center coordinate space. Due to the area-preserving symmetry of the guiding center plane (the continuum version of relabeling the discrete electrons), we will have another constraint derived from the conserved quantity associated with the area-preserving symmetry. This adds a Lagrange multiplier term that helps us complete the Abelian Chern-Simons Lagrangian.  The resulting effective Chern-Simons theory is exactly identical to the first-order expansion in the noncommutative parameter of the noncommutative Chern-Simons theory. This suggests us that the noncommutative Chern-Simons theory may be a better description for the physics in the lowest Landau level {than the commutative Chern-Simons theory}.

It is natural to relate the above mentioned position fluctuations of electrons to the metric fluctuation $\delta g^{ab}({\bf x},t)$ around the guiding center metric $g_0^{ab}$ in the Haldane's proposal. Our observation is that if we identify the spatial gauge fluctuation with the {\it zweibein} fluctuations of the metric $\delta g^{ab}$, the noncommutative Chern-Simons Lagrangian turns out to be a theory of two-dimensional geometry with a flat time. The area-preserving constraint becomes a solution to the field equations for the emergent geometry. Here the emergent geometry is a dynamic one -- due to the collective position fluctuations relative to the guiding center -- not an external one as introduced by Wen and Zee \cite{wenzee}.
Furthermore the emergent quantum geometry inspires the identification in our unified approach of two important topological indices for the FQHE, which were proposed before separately in Refs. \cite{wenzee,haldane,haldane2} : (i) the shift which was thought of as an 'orbital angular momentum' of the guiding center and is related to the filling factor; (ii) the guiding center spin, or the 'spin' of the guiding center, which is a topological index characterizing the Laughlin state besides the filling factor.
We can also generalize our results to hierarchical FQH states, giving a geometric description for the gauge fluctuations in the K-matrix Chern-Simons theory.

This paper is organized as follows, in the following section we will first review the guiding center description of FQHE. Then we show the emergence of the Chern-Simons action as the dynamic degrees of freedom of guiding center at the lowest Landau level {in two ways, namely, by means of both Dirac bracket and Moyal $*$-product}. 
In Sec. III, we will {use the Dirac bracket method to }discuss the emergent geometry arising from the noncommutative Chern-Simons gauge theory, and we will generalize our results to other FQH states in K-matrix formalism in Sec. IV. {In Appendix A, we present a short introduction to noncommutative geometry. The quantum geometric interpretation of the emergent noncommutative Chern-Simons theory of the lowest Landau level is discussed in Appendix B.}

\section{Emergence of Chern-Simons gauge theory}
\subsection{Review of guiding center description of FQHE}
The two-dimensional interacting electron gas in a perpendicular magnetic field is described by the Lagrangian
\begin{eqnarray}
L=\sum_{i=1}^Ng_G^{ab} [\frac{m_b}2\dot x_{ia}\dot x_{ib}+\dot x_{ia} A_{ib}]-\sum_{i<j}\frac{1}{\varepsilon
[g_c^{ab}x_{ija}x_{ijb}]^{1/2}}.  \label{1}
\end{eqnarray}
Here we used Einstein's summation convention for $a,b=1,2$ labeling two-dimensional coordinates; $x_{ia}$ is the position and $\dot x_{ia}$ the velocity, respectively, of the $i$ th electron, and $x_{ija}=x_{ia}-x_{ja}$. We have assumed that the effective band mass tensor is of the form $m_b g_G^{ab}$, which defines both the band mass $m_b$ and the "Galilean metric" $g_G^{ab}$ with the condition $\det(g_G)=1$. The metric $g_c^{ab}$ is the Coulomb metric arising from the small momentum behavior of the Coulomb potential for electron-electron interactions, if there is microscopic anisotropy (due to, e.g., the environment of the two-dimensional electron gas). Both $g_c$ is assumed unimodular:  $\det(g_c)=1$; this fixes the value of the dielectric constant $\varepsilon$. Introducing the two metrics allows us to deal with the more general cases with spatial anisotropy. (The usual Laughlin wave function approach has assumed the rotational symmetry or spatial isotropy.) $A_{ia}$ is the vector potential of the external uniform magnetic field  ${\bf B}$. In the symmetric gauge, $A_a=B\epsilon_{ab}x^b/2$. For simplicity, we take the units $e=\hbar=c=l_B=1$ ($l_B$ being the magnetic length). The kinematic momenta $\pi_a=-i\partial_a+A_a$ are noncommutative: $[\pi_{ia},\pi_{jb}]=i\epsilon_{ab}\delta_{ij}$. Define the guiding center operators as:
\begin{equation}
\hat Y_{ia}=x_a-\epsilon_{ab}\pi^b_i.\label{206}
\end{equation}
The physical meaning of $\hat{Y}_{ia}$ is that it represents the position of the center, the guiding center, of the cyclotron motion of the $i$-th electron. Notice that the guiding center operators commute with the kinematic momenta $\pi_a$:  $[Y_{jb},\pi_{ia}]=0$. This implies that the guiding center coordinates describe the degrees of freedom (for a single electron) within the lowest Landau level.
In the case when $g_c\ne g_G$, there will be an emergent metric $g_0$ that minimizes the correlation energy\cite{haldane}. The $g_0$ is called the guiding center metric, since they are defined for the degrees of freedom in the LLL \cite{haldane}. Several examples with $g_c\ne g_G$ were studied, and the $g_0$ has been worked out explicitly in refs. \cite{ex,ex1,ex2,ex3,ex4}.

Classically, the position of a two-dimensional electron in a perpendicular uniform magnetic field can be decomposed into that of its guiding center plus a cyclotron motion around the guiding center. We use $y_{ia}$ to denote the guiding center coordinates;  $\hat Y_{ia}$ are the corresponding quantum operators. The electron position then can be decomposed into
\begin{equation}
x_{ia}=y_{ia}+\delta x_{ia}(t),
\end{equation}
where $\delta x_{ia}(t)$ is the electron position deviation from $y_{ia}$, which are { the dynamic degrees of freedom} and have both the kinematic and collective origins. In the lowest Landau level the kinematic part is frozen since the kinematic energy part is projected out, while the collective modes survive and depend only on $y$, the guiding center. We write the collective degrees of freedom as
\begin{eqnarray}
 x_{ia}=y_{ia}+\theta\epsilon_{ab}a_i^b(y,t)\label{2},
 \end{eqnarray}
where $a^b$ is actually a Chern-Simons gauge potential as will be shown below. The constant $\theta$ is taken to be $\theta=1/\nu=m$ for later convenience.

Define the {\it zweibein} $e_{0\alpha}^a$ by $g_{0ab}= e_{0\alpha a}e_{0b}^\alpha$,
with $\alpha=1,2$ the frame index. Since $g_0$ is symmetric and unimodular, the vector $x_{ia}$ may be written as
\begin{equation}
x_{ia}=x_{i\alpha} e^\alpha_{0a}=y_{i\alpha}e^\alpha_{ia}.
\end{equation}
$e^\alpha_{ia}$ looks like a local "{\it zweibein}". However, it is difficult to locally define a {\it zweibein} or a metric in a discrete system. This motivates us to study an emergent geometry in the { continuum} effective theory which is discussed in detail in Sec. III.

\subsection{The zero band mass limit and Dirac brackets}

We now exhibit the noncommutativity of the coordinates in the lowest Landau level. In the zero band mass limit $m_b\to 0$,  the kinetic energy vanishes, which projects the system to the lowest Landau level. The degrees of freedom related to $\pi_{ia}$, the "left-handed ones" \cite{haldane}, are frozen in the Hamiltonian.  Here, we would like to emphasize that the dynamic degrees of freedom are not totally congealed in the sense of a topological quantum mechanics which involves in the collective dynamic degrees of freedom,  i.e., replacing the  free Lagrangian in Eq. (\ref{1}) by{
\begin{eqnarray}
L_0=\sum_{i=1}^N\dot{x}_i^aA_{ia}=\frac{1}2\sum_{i=1}^N \epsilon^{ab}x_{ia}\dot x_{ib},\label{L0}
\end{eqnarray}}
which describes the Chern-Simons quantum mechanics of the electrons subject to the second class constraints\cite{jac}{
\begin{eqnarray}
\pi_{ia}=P_{ia}-A_{ia}\sim 0.\label{constraint}
\end{eqnarray}}
Here $\sim$ stands for being set to zero { merely after} calculating the commutation
relations  $[\pi_{ia},\pi_{jb}]=i\delta_{ij}\epsilon_{ab}$ \cite{jac}.
Owing to the second class constraints, quantization should be carried out by using Dirac brackets \cite{dirac, dirac2, dirac3},
\begin{equation}
\{O_{1i},O_{2j}\}_D=[O_{1i},O_{2j}]-[O_{1i},\pi_{ka}]C^{ab}_{kl}[\pi_{lb},O_{2j}], \label{00}
\end{equation}
where $C^{ab}_{kl}$ are defined by $C^{ab}_{kl}[\pi_{lb},\pi_{nc}]=\delta^a_{c}\delta_{kn}$, i.e., $C^{ab}=-i\epsilon^{ab}\delta_{kl}$. 
{Due to the noncommutative nature of the operators in quantum mechanics, here we shall emphasize that after quantization, the Dirac bracket method could suffer from the ordering of operators\cite{dirac3}. With the $C^{ab}_{kl}$ being constant in our special case, this ordering problem may disappear, though the application of a Dirac bracket in quantum theory, in general, is still delicate. Keeping this in mind, we have}
\begin{equation}
\{x_{ia},x_{jb}\}_D=i\epsilon_{ab}\delta_{ij}, \label{201}
\end{equation}
which reflects the noncommutativity of electron positions in the lowest Landau level. This noncommutativity tells us that the geometry of the lowest Landau level is better described as a noncommutative geometry and there are some collective modes. The existence of the collective modes is due to the nonzero $[x_{ia},x_{jb}]_D$; otherwise, the Dirac bracket would be zero. Since $[\pi_{ia},\hat Y_{jb}]=0$, Dirac brackets of $\hat Y_{ia}$'s are the same as the usual commutators:
\begin{equation}
\{\hat Y_{ia},\hat Y_{jb}\}_D = [\hat Y_{ia},\hat Y_{jb}]=-i\epsilon_{ab}\delta_{ij}. \label{202}
\end{equation}
{From Eq. (\ref{201}) and Eq. (\ref{202}), we show that not only the electron position space but also the guiding center space is noncommutative in the lowest Landau level. This suggests us that we shall use a formalism that deals with coordinate noncommutativity. We have two choices, one is the Dirac bracket formalism and the other is the Moyal $*$ product\cite{jemo}. In order to keep the consistency of logic and to keep our derivation  fundamental, we shall continue using the Dirac bracket formalism. The relationship between the Dirac bracket formalism and the Moyal $*$ product method will be discussed at the end of this section. We will show that these two methods are parallel and consistent with each other.  }

\subsection{Continuum limit and Chern-Simons gauge theory}

{In order to discuss the emergent geometry in the lowest Landau level, we first consider the continuum limit of the system. Starting from the identity for an arbitrary function of $x$,{
\begin{equation}
\sum_i f(x_i)=\int dx \sum_i\delta(x-x_i)f(x)
=\int dx \rho(x;x_i)f(x),\label{d3}
\end{equation}}
where $\rho(x;x_i)\equiv  \sum_i\delta(x-x_i)$. 
Because we will discuss the physics in the guiding center space, we transform the above equation from the position space $x$ to the guiding center space $y$, and obtain 
\begin{eqnarray}
\sum_i f(x_i) &=&\int dx \sum_i\delta(x-x_i)f(x)\nonumber\\
&=&\int dy \rho_0 (y;y_i) f(x(y)).
\end{eqnarray}
where $y_{ia}$ are the guiding center coordinates, given by (\ref{206}),
\begin{equation}
 y_{ia}=x_{ia}-\epsilon_{ab}\pi^b_i,
\end{equation}	
and {
\begin{equation}
\rho_0(y;y_i)= 
\rho(x(y;y_i))|\det(\frac{\partial x}{\partial y})|,\label{d6}\\ 
\end{equation} }
Thus the continuum form of the Lagrangian (\ref{L0}) in guiding center space is given by  {
\begin{eqnarray}
L_0&=&\sum_{i=1}^N\dot{x}_i^aA_{ia}=\frac{1}2\sum_{i=1}^N \epsilon^{ab}x_{ia}\dot x_{ib}\nonumber\\
&=&\int dx \rho(x) \frac{1}{2}\epsilon^{ab} x_a\dot{x}_b\nonumber\\
&=&\int dy \rho_0(y) \frac{1}{2}\epsilon^{ab}x_a(y,t)\dot{x}_b(y,t), \label{d4}
\end{eqnarray}
where for later convenience we have abbreviated $\rho(x;x_i)$ as $\rho(x)$. }

In the following we only consider the situation that { there is no vortex excitation.} With this assumption, we know that the guiding center density, $\rho_0 = \frac{\nu}{2\pi l_B^2}$, is always uniform. The (quantum) incompressibility of the FQH liquid implies that the residual dynamic degrees of freedom are totally determined by the collective modes, so the requirement that the $x$ fields are the functions of the guiding center coordinates $y$ only is justified. 

We follow a similar treatment in Ref. \cite{nccs}. Let us consider the area-preserving transformation in the guiding center coordinates, the $x$ field behaves like a scalar field and the Lagrangian (\ref{d4}) is invariant under this area-preserving transformation. This area-preserving symmetry comes from relabeling the electrons in the discrete version\cite{nccs}.{ Also this symmetry is a gauge symmetry, whose physical origin lies in the lowest Landau level projection, as we now argue. Classically, if one needs to determine the position of the guiding center, one will need to know both the position and the momentum of the electron, as in the definition of guiding center (\ref{206}). But after the lowest Landau level projection, the kinematic energy part is suppressed. Therefore we lose the information about the momentum of electron. The ambiguity in determining the position of the guiding center is the physical origin of this emergent gauge symmetry. }

Under an infinitesimal area-preserving transformation,{
\begin{equation}
y'_a=y_a+\epsilon_{ab}\frac{\partial \Delta(y)}{\partial y_b},
\end{equation}}
where $\Delta$ is an arbitrary gauge function. Then{
\begin{equation}
\delta x_a=\epsilon_{cd}\frac{\partial x_a}{\partial y_c}\frac{\partial \Delta}{\partial y_d}.
\end{equation}}
According to Noether's theorem, there is a conserved quantity $\Theta$ associated with this area-preserving symmetry,{
\begin{equation}
\Theta= \frac{\delta L_0}{\delta \dot{x}_a}\delta x_a=\frac{\rho_0}{2}\int d^2 y\epsilon_{ab}\epsilon^{cd}x_c\frac{\partial x_d}{\partial y_a}\frac{\partial \Delta}{\partial y_b}.
\end{equation}}
Since $\Theta$ is conserved, i. e., $\dot{\Theta}=0$, and $\Delta$ is an arbitrary function, we can conclude that (after integrating by parts),{
\begin{eqnarray}
&&\frac{d}{dt}(\frac{1}{2}\frac{\partial}{\partial y_b}(\epsilon_{ab}\epsilon^{cd} x_c\frac{\partial x_d}{\partial y_a}))=\frac{d}{dt}(\frac{1}{2}\epsilon_{ab}\epsilon^{cd}\frac{\partial x_d}{\partial y_a}\frac{\partial x_c}{\partial y_b})\nonumber\\
&&=\frac{d}{dt}(\det(\frac{\partial x}{\partial y}))=0.
\end{eqnarray}}
This tells us that the Jacobian between $x$ and $y$ is independent of time. In the absence of vortices, this Jacobian is chosen as unity\cite{nccs}, i.e.,
\begin{equation}
\det\left( \frac{\partial x}{\partial y}\right) =1. \label{area}
\end{equation}
From Eq. (\ref{d6}), we conclude that by choosing the determinant $\det(\partial x/\partial y)=1$, the electron density $\rho(x)$ equals the guiding center density $\rho_0(y)$ and becomes a constant after the lowest Landau level projection when there are no vortices, which means that the electrons form an incompressible fluid in both guiding center space and the electron coordinate space. 

Another ingredient we need to discuss is the generalization of Dirac brackets in their continuum form, if we project the system to the lowest Landau level. In the continuum limit, Eq. (\ref{constraint}) becomes,
\begin{equation}
\Pi^a(x(y))=P^a(x(y))-A^a(x(y))\sim 0,
\end{equation}
where the notation $f(x(y))$ means that any function of the electron position space $x$ can be written as a composite function of the guiding center space $y$ and because $\det(\partial x/\partial y)=1$, the coordinate transformation form $x$ to $y$ is one to one. The commutation relation of the $\Pi$s is
\begin{equation}
[\Pi^a(y),\Pi^b(y')]=i\epsilon^{ab}\delta(y-y'),\label{d5}
\end{equation}
where we have assumed the canonical commutation relation:
\begin{equation}
[P^a(y), X^b(y')]=i\delta^{ab}\delta(y-y').
\end{equation}
Here we shall keep in mind that the $X^a(y)$ is a field and $P^a(y)$ is its conjugate momentum field. 

In order to define the Dirac bracket in the continuum limit, we shall first solve the following equation,
\begin{equation}
\int dy' [\Pi^a(y),\Pi^c(y')]C_{cb}(y',y'')=\delta^{a}_b\delta(y-y'').
\end{equation}
Because of Eq. (\ref{d5}), it is easy to find that
\begin{equation}
C_{ab}(y,y')=-i\epsilon_{ab}\delta(y-y'),
\end{equation} 
which is also consistent with the discrete version, namely, $C^{ab}_{kl}=-i\epsilon^{ab}\delta_{kl}$. Then we can define the Dirac bracket in the continuum limit,
\begin{eqnarray}
&&\{O_1(y_1), O_2(y_2)\}_D=[O_1(y_1), O_2(y_2)]\nonumber\\
&&-\int dydy' [O_1(y_1),\Pi_a(y)]C^{ab}(y,y')[\Pi_b(y'),O_2(y_2)].\nonumber\\
\end{eqnarray}
How to deal with $[O_1(y_1),\Pi_a(y)]$ is a little bit tricky. In the following we will only encounter fields that are functions of the electron position field $X$; therefore, we will only consider the $O_1$ field as a function of $X$, then
\begin{equation}
[O_1(X_1),\Pi_a(x)]=i\frac{\partial O_1(X)}{\partial X^a}\delta(X_1-x).
\end{equation}
Now we shall regard the field $X^a$ in the $\partial X^a$ as the electron coordinate $x^a$, i.e., $\partial x^a$. Then,
\begin{eqnarray}
&&\{O_1(Y_1),O_2(Y_2)\}_D\nonumber\\
&=&\delta(Y_1-Y_2)i\epsilon^{ab}\frac{\partial O_1(x(Y_1))}{\partial x^a}\frac{\partial O_2(x(Y_1))}{\partial x^b}\nonumber\\
&=&\delta(Y_1-Y_2)\det(\frac{\partial x}{\partial y})i\epsilon^{ab}\frac{\partial O_1(Y_1)}{\partial y^a}\frac{\partial O_2(Y_1)}{\partial y^b}\nonumber\\
&=&\delta(Y_1-Y_2)i\epsilon^{ab}\frac{\partial O_1(Y_1)}{\partial y^a}\frac{\partial O_2(Y_1)}{\partial y^b}.\label{defdirac}
\end{eqnarray}
where we have used the ordinary commutator $[O_1(X(Y_1)),O_2(X(Y_2))]=0$ and $\det{\partial x/\partial y}=1$. In the following, Eq. (\ref{defdirac}) will be the definition of the Dirac bracket in the continuum limit. {From Eq. (\ref{defdirac}), we can directly calculate that 
\begin{equation}
\{x_a(y'), x_b(y'')\}_D=i\epsilon_{ab}\delta(y'-y''),
\end{equation}
which is a natural generalization of the discrete version (\ref{201}).}

With all the preparations above, we can now show the emergence of a noncommutative Chern-Simons theory which describes the collective behavior of the system after the lowest Landau level projection. Recall that Eq. (\ref{area}) is actually a constraint imposed by the area-preserving symmetry; therefore, we shall introduce a Lagrange multiplier $a_0$ to add the constraint (\ref{area}) into (\ref{d4})},  
then the Lagrangian (\ref{d4}) becomes
\begin{equation}
L_0=\frac{\rho_0}{2}\epsilon^{ab}\int_{R^2} d^2
y[(\dot{x}_a(y)-\frac{1}{2\pi\rho_0}\{x_a,a_0\}_P)x_b+\frac{\epsilon_{ab}}{2\pi\rho_0}a_0], \label{208}
\end{equation}
where we used the notation $[\cdot,\cdot]_P$ for the "Poisson" bracket\cite{nccs}, which is defined as
\begin{equation}
\{F(y),G(y)\}_P=\epsilon^{ab}\partial_a F\partial_b G .
\end{equation}
One can verify the area-preserving condition (\ref{area}) by calculating the equation of motion of $a_0$.  The "Poisson" bracket is an analog of the Poisson bracket defined in the phase space. Because now we are dealing with noncommutative geometry, here we can think of the guiding center coordinates $\{y_a\}$ as some kind of phase space. We shall emphasize the difference between our treatment and the treatment in \cite{nccs}. The key difference of the underlining physics is that the area-preserving symmetry here is fulfilled by the diffeomorphisms in the guiding center description, while the diffeomorphisms in \cite{nccs} are explained as the Eulerian description of a fluid. 

Substituting the continuum version of Eq. (\ref{2}), i.e., $x_a=y_a+\theta\epsilon_{ab}a^b(y,t)$, into Eq. (\ref{208}) {and using the Dirac bracket (\ref{defdirac}) instead of the Poisson bracket}, we have,{
\begin{eqnarray}
L_0&=&\frac{\rho_0}{2}\int _{S^2}d^2y(\theta^2\epsilon^{cd}a_0\partial_c a_d+\theta^2 \epsilon^{cd}a_c\partial_d a_0\nonumber\\&-&\theta^2\epsilon^{cd}a_c\partial_0 a_d+\theta^3\epsilon^{ab}\epsilon^{cd}a_0\partial_c a_a \partial_d a_b )\nonumber\\
&=&\frac{1}{4\pi \nu}\int _{S^2}d^2y\epsilon^{\mu\nu\rho}(a_\mu\partial_\nu
a_\rho+\frac{\theta}{3}\epsilon^{ab}a_\mu\partial_a
a_\nu\partial_b a_\rho) .\nonumber\\ 
\label{9}
\end{eqnarray}}
Here $\mu=0,1,2$ and the Lagrange multiplier $a_0$ is identified as the zero component of Chern-Simons potential while the position fluctuations as the spatial components of the gauge potential. Although we added the the Lagrange multiplier $a_0$ by hand, after a careful calculation, the final Lagrangian (\ref{9}) is symmetric in $a_0$, $a_1$, and $a_2$. One can verify the correctness of (\ref{9}) quickly by calculating the equation of motion of $a_0$, which turns out to be the constraint equation (\ref{area}). $R^2$ is compactified, for convenience, to a sphere $S^2$ because the gauge potential $a_a=0$ at $|{\bf y}|=\infty$. This is required by vanishing of the position fluctuations at the infinity. This also gives the gauge invariance of $L_0$ \cite{nccs}.

{
Therefore we have shown the emergence of the Chern-Simons theory, i.e.,
\begin{eqnarray}
L_0=\frac{1}{4\pi \nu}\int_{S^2}d^2y \epsilon^{\mu\nu\rho}\left(a_\mu\partial_\nu a_\rho+\frac{\theta}{3}\epsilon^{ab}a_\mu \partial_a a_\nu \partial_b a_\rho\right), ~~\label{l00}
\end{eqnarray}
with the gauge transformation 
\begin{equation}
\delta a_a=\frac{\partial \Delta}{\partial y^a} + \theta\{a_a,\Delta\}_{D},\label{34v2}
\end{equation}
and the constraint equation becomes
\begin{equation}
\epsilon^{ab}\left(\frac{\partial a_b}{\partial y^a}-\frac{\theta}{2}\{a_a,a_b\}_{D}\right)=0.\label{209}
\end{equation}
{The gauge transformation (\ref{34v2}) and the constraint equation (\ref{209}) will remind us of the noncommutative Abelian Chern-Simons theory. The results here shed some light on the equivalence between the method of the Dirac bracket and the noncommutative $*$ product. As mentioned before, we will now discuss the relationship between them in the following subsection. }

\subsection{Relationship with noncommutative geometry}

{Now we use the language and techniques of noncommutative geometry for the problem of many electrons in the lowest Landau level. {We will derive results similar  to Eq. (\ref{l00}), (\ref{34v2}), and (\ref{209}) from the perspective of noncommutative geometry, which suggests the consistency between the Dirac bracket and the Moyal $*$-product.} For our purpose, it is more convenient to examine a continuum field theory instead of the $N$-body quantum mechanics. In the latter framework, the many-body wave function lives in a higher-dimensional space with $2N$ coordinates. However, in the framework of a field theory, one is able to use fields living in the space of only a pair of coordinates, to describe the collective behavior of a many-body system. { Therefore} noncommutative geometry becomes suitable for making the transition from a discrete particle formulation to field theory{(a short introduction to noncommutative geometry is presented in  Appendix\ref{appen1}).}  }

{Thus, we proceed to describe the degrees of freedom of the system of electrons in the lowest Landau level by introducing the electron "position fields" $x_a(y,t)$ ($a$=1,2) living in the noncommutative plane $R^2$. This plane has two continuous coordinates $y=(y_1,y_2)$, which we identify to be the guiding center coordinates in the lowest Landau level. For the fields $x_a(y,t)$, the coordinates $(y_1,y_2)$ play a role similar to the particle label $i$ in the main text. We will argue that a natural Lagrangian of the theory is given by      
\begin{eqnarray}
L_0&=&\frac{1}2\int _{R^2}d^2y\rho_0 \epsilon^{ab}x_a(y,t)*\dot x_b(y,t),\label{alaggg}\\
&\sim&\frac{1}2\int _{R^2}d^2y\rho_0 \left[\epsilon^{ab}x_a(y,t)\dot x_b(y,t)+\frac{i}{2}\theta\frac{d}{dt}\det(\frac{\partial x}{\partial y})\right],\nonumber \\\label{alag}
\end{eqnarray}
with the $*$-product defined by Eq. (\ref{204}) for functions or fields dependent on $y_a$ $(a,b=1,2)$: 
\begin{eqnarray}
f(y)*g(y)& = & \exp\left[\frac{i}{2}\theta
\epsilon^{ab}\frac{\partial}{\partial\xi^a}
\frac{\partial}{\partial \eta^b}\right] \nonumber\\
&&f(y+\xi)g(y+\eta)|_{\xi=\eta=0}.
\end{eqnarray} 
We need the $*$-product here to enforce the 
noncommutativity of the guiding center coordinates after the lowest Landau level projection. As one will see, the $*$-product of two functions at the same $y$ is sufficient for our following treatments for the action and equations of motion of an emergent noncommutative Chern-Simons gauge theory. 

{If we expand the $*$-product to the zeroth order and choose the density function to have the form $\rho_0(y;y_i)=\sum_i \delta^2(y-y_i)$ (with $i$ the particle label), then Eq. (\ref{alaggg}) will reduce to Eq. (\ref{L0}). This tells us that the Lagrangian (\ref{alaggg}) is a reasonable continuum limit of the discrete Lagrangian (\ref{L0}), which incorporates as well the noncommutativity of electron positions in the lowest Landau level as well.}} Equation (\ref{alag}) is the first order expansion in the noncommutative parameter $\theta$ of the $*$-product. 

{We will assume that there are no vortices. This incompressible FQH fluid implies that the collective modes are the residual dynamic degrees of freedom after the lowest Landau level projection.} 

{Now let us consider the area-preserving symmetry}. Under an infinitesimal area transformation, 
\begin{equation} 
y'_a=y_a+\epsilon_{ab}\frac{\partial \Delta(y)}{\partial y_b},
\end{equation}
where $\Delta$ is an arbitrary gauge function. Then $x'$ transforms as 
\begin{eqnarray}
\delta x_a&=&\frac{\partial x_a}{\partial y_c}*(\epsilon_{cd}\frac{\partial \Delta}{\partial y_d})\nonumber\\
&\sim&\epsilon_{cd}\frac{\partial x_a}{\partial y_c}\frac{\partial \Delta}{\partial y_d}+\frac{i}{2}\theta\epsilon^{cd}\epsilon^{ef}\frac{\partial^2 x_a}{\partial y^c \partial y^e}\frac{\partial^2 \Delta}{\partial y^d \partial y^f}.\label{207}
\end{eqnarray}
 {Although defining the area-preserving diffeomorphism in noncommutative space conceptually is delicate, we can consider power expanding $\theta$ terms to the first order and study the corrections due to the infinitesimal area-preserving transformation.} Therefore, let us focus on the linear-order term in (\ref{207}), i.e., the first term; we will deal with the nonlinear contribution later. The conserved quantity $\Theta$ associated with this area-preserving symmetry is,
\begin{equation}
\Theta= \frac{\delta L_0}{\delta \dot{x}_a}\delta x_a=\frac{\rho_0}{2}\int d^2 y\epsilon^{dc}x_d\epsilon_{ab}\frac{\partial x_c}{\partial y_a}\frac{\partial \Delta}{\partial y_b}.
\end{equation}
Since $\Theta$ is conserved, i. e.,
\begin{equation}
\frac{d}{dt}\left( \frac{\rho_0}{2}\int d^2 y\epsilon^{dc}x_d\epsilon_{ab}\frac{\partial x_c}{\partial y_a}\frac{\partial \Delta}{\partial y_b} \right)=0,
\end{equation}
 and $\Delta$ is an arbitrary function, which means that,
\begin{eqnarray}
&&\frac{d}{dt}(\frac{1}{2}\frac{\partial}{\partial y_b}(\epsilon^{cd}\epsilon_{ab} x_c\frac{\partial x_d}{\partial y_a}))=\frac{d}{dt}(\frac{1}{2}\epsilon^{cd}\epsilon_{ab}\frac{\partial x_c}{\partial y_a}\frac{\partial x_d}{\partial y_b})\nonumber\\
&&=\frac{d}{dt}(\det(\frac{\partial x}{\partial y}))=0.
\end{eqnarray}
Then we can choose this Jacobian to be unity, which is a new constraint for the system,
\begin{equation}
\det(\frac{\partial x}{\partial y})=1 .  \label{aarea}
\end{equation}
{Requiring the Jacobian to be unity is a reasonable constraint in our treatment since, as one can easily check, it leads to  
\begin{equation}
[x_a(y),x_b(y)]_*=i\theta \epsilon_{ab},
\end{equation}
to the first order in $\theta$.}

We introduce a Lagrange multiplier $a_0$ to add the constraint
(\ref{aarea}) into (\ref{alag}), then the Lagrangian (\ref{alag}) becomes
\begin{equation}
L_0=\frac{\rho_0}{2}\epsilon^{ab}\int_{R^2} d^2
y[(\dot{x}_a(y)-\frac{1}{2\pi\rho_0}\{x_a,a_0\}_P)x_b+\frac{\epsilon_{ab}}{2\pi\rho_0}a_0], \label{a208}
\end{equation}
with $\{\cdot,\cdot\}_P$ stands for the "Poisson" bracket\cite{nccs}, which is defined as
\begin{equation}
\{F(y),G(y)\}_P=\epsilon^{ab}\partial_a F\partial_b G .
\end{equation}
{ By varying $a_0$, one can check the area-preserving condition (\ref{aarea}). Since we are using the language of noncommutative geometry, we can regard the guiding center space  $\{y_a\}$ as phase space, then, the "Poisson" bracket used here becomes an analog of the Poisson bracket defined in classical mechanics.}

Substituting the continuum version of Eq. (\ref{2}), i.e., $x_a=y_a+\theta\epsilon_{ab}a^b(y,t)$, into Eq. (\ref{a208}), we have,
\begin{eqnarray}
L_=\frac{1}{4\pi \nu}\int _{S^2}d^2y\epsilon^{\mu\nu\rho}(a_\mu\partial_\nu
a_\rho+\frac{\theta}{3}\epsilon^{ab}a_\mu\partial_a
a_\nu\partial_b a_\rho) .\nonumber\\ \label{a9}
\end{eqnarray}
Here $\mu=0,1,2$ and the Lagrange multiplier $a_0$ is identified as the zero-component of Chern-Simons potential while the position fluctuations as the spatial components of the gauge potential. 
Because we have only considered the linear contribution of (\ref{207}) to the constraint equation (\ref{aarea}), we shall propose to replace the "Poisson" bracket in (\ref{a208}) with the noncommutative bracket to incorporate the higher order contributions to (\ref{207}). Then we come up with the noncommutative Chern-Simons Lagrangian,
\begin{eqnarray}
L_0=\frac{1}{4\pi \nu}\int_{S^2}d^2y \epsilon^{\mu\nu\rho}\left(a_\mu*\partial_\nu a_\rho+\frac{2i}{3}a_\mu * a_\nu * a_\rho\right). ~~\label{al00}
\end{eqnarray}
If we expand the Lagrangian (\ref{al00}) to the first order in $\theta$, it is identical to Eq. (\ref{a9}). The corresponding noncommutative gauge transformation becomes
\begin{equation}
\delta a_a=\frac{\partial \Delta}{\partial y^a} + \theta[a_a,\Delta]_{*},\label{a208v2}
\end{equation}
and the constraint equation becomes
\begin{equation}
\epsilon^{ab}\left(\frac{\partial a_b}{\partial y^a}-\frac{\theta}{2}[a_a,a_b]_{*}\right)=0.\label{a209}
\end{equation}
{By comparing the noncommutative Lagrangian (\ref{a9}), the noncommutative gauge transformation (\ref{a208v2}), and the constraint equation (\ref{a209}) with Eq. ({\ref{l00}}), (\ref{34v2}), and (\ref{209}), we can find that the results are consistent and parallel with each other in both formalisms. As a final remark of this section, to facilitate the comparison with the usual (commutative) field theory in the limit $\theta \to 0$, we have used an expansion of the $*$-product (\ref{204}) up to first order in the noncommutativity parameter $\theta$, though the $\theta$ expansion is formal and its convergence is not evident.}

\section{Emergent geometry from Chern-Simons gauge theory}
\subsection{Geometric description for Chern-Simons theory}
Now we show that a quantum geometry emerges from the gauge theory  (\ref{l00}). To this end, we define the fluctuating unimodular metric $g_{ab}({\bf y},t)=g_{0ab}+\delta g_{ab}=e^\alpha_ae_{\alpha b}$, where the {\it zweibein} is parametrized by the gauge field, i.e.,
 \begin{eqnarray}
 e^\alpha_a=(e^\alpha_{0a}+\sqrt{2\theta/g_{012}}\epsilon_{ab}\delta^{bc}e^{\alpha}_{0c}a_c)/\sqrt{N_g}, \label{zwei}
 \end{eqnarray}
 which is the continuous version of (\ref{2}) with the normalization factor $N_g=1+2\theta a_1a_2/g_{012}$ such that  $\det g^{ab}=\det g_{ab}=1$. Here we have used a modified Einstein summation convention; i.e., we also sum over the situation when there are one up dummy index and two down dummy indices [e.g., the index $c$ in (\ref{zwei})].  Since $g_{ab}$ is a symmetric and unimodular tensor, the gauge field can also be expressed by the {\it zweibein} with no redundancy, i. e., the number of degree of freedom is the same. Let us multiply $e_{0d\alpha}$ to both sides of (\ref{zwei}), and pick out the cases of $a=1,d=1$ and $a=2,d=2$, respectively:
    \begin{eqnarray*}
    e_1^\alpha e_{01\alpha}&=&(g_{011}+\sqrt{2\theta g_{012}}\epsilon_{12}a_2)/\sqrt{N_g}, \\
    e_2^\alpha e_{02\alpha}&=&(g_{022}+\sqrt{2\theta
    g_{012}}\epsilon_{21}a_1)/\sqrt{N_g}.
    \end{eqnarray*}
 Therefore we can write the results in a more compact form,
 \begin{eqnarray}
 a_b=\epsilon_{bc}\delta^{cd}(g_{0dd}-{\cal G}_{dd} {\cal C}(e))/\sqrt{2\theta g_{012}},
 \end{eqnarray}
 where ${\cal G}_{ab}=e^\alpha_{0a}e_{\alpha b}$ and ${\cal C}(e)$ is the positive solution of the equation
 $({\cal G}_{11}{\cal G}_{22}+g_{012}^2){\cal C}^2-({\cal G}_{22}g_{011}+{\cal G}_{11}g_{022}){\cal C}-1=0$. (In fact, ${\cal C}=\sqrt{N_g}$ in the $a_b$ parametrization.)

The  spin connection corresponding to $e$ is given by
\begin{eqnarray}
\Omega_t&=&\epsilon^{ab}e^{\alpha}_a\partial_te_{b\alpha }/2,\nonumber\\
\Omega_i&=&\epsilon^{ab}e^{\alpha}_a D_ie_{b\alpha}/2=(\epsilon^{ab}e_a^\alpha\partial_i e_{b\alpha}-\epsilon^{ab}\partial_a g_{bi})/2,\label{CD}
\end{eqnarray}
where $D_a$ is the covariant derivative with the Levi-Civita  connection $\Gamma_{ab}^c$. The corresponding Gauss curvature $R$ is defined as{
    \begin{equation}
    R\equiv \epsilon^{ab}\partial_a\Omega_b=\frac{1}{2}\epsilon^{ab}\epsilon^{cd}(\partial_ae^\alpha_c\partial_be_{d\alpha}-\partial_c\partial_ag_{db}) .
    \end{equation}}
Then we can rewrite the Chern-Simons Lagrangian (\ref{9}), in terms of the spin connection $\Omega_t$, $\Omega_i$ and Gauss curvature $R$, as
\begin{eqnarray}
L_0&=&\frac{1}{4\pi \nu}\int_{S^2} d^2y[a_0\sqrt{\frac{2}{\theta g_{012}}}\partial^a({\cal G}_{aa}{\cal C})\nonumber\\
&-&\frac{\Omega_t{\cal C}^2}{\theta}-(\epsilon^{ab} )^2 g_{0aa}\frac{{\cal C}}{\theta g_{012}}\partial_t{\cal G}_{bb}\nonumber\\
&+&J(e)(R+\frac{1}{2}\epsilon^{ab}\epsilon^{cd}\partial_a\partial_c g_{bd}) a_0],~~\label{lo}
\end{eqnarray}
 where
 \begin{eqnarray}
 J(e)=N_g^2 = (1+2\theta a_1a_2/g_{012})^2.\label{f}
 \end{eqnarray}
If we expand $J(e)=1+O(a)$, the coefficient of the Gauss curvature in Eq. (\ref{lo}) is $J(e)/4\pi\nu=1/4\pi\nu+O(a)$. To the zeroth order of the Chern-Simons field, this gives a correct shift ${\cal S}=1/\nu=\theta=m$ for the $\nu=1/m$ Laughlin state on a sphere \cite{wenzee}. $L_m={\cal S}/2$  is a topological invariant and is identical as the guiding center orbital momentum\cite{wenzee}. And also we can rewrite the $L_0$ (\ref{lo}) in terms of the gauge field $a_\mu$,
   \begin{eqnarray}
   L_0&=&\int_{S^2} d^2y\biggl[\frac{1}{4\pi \nu}\epsilon^{\mu\nu\rho}a_\mu\partial_\nu a_\rho
   +\frac{\theta^2}{2\pi}(\frac{1}{2\theta}\nonumber\\
   &+&\frac{1}{g_{012}}a_1a_2)\epsilon^{ab}\Omega_a\partial_b a_0+\cdots\biggr],\label{loother}
    \end{eqnarray}
where $\cdots$ are terms containing the Levi-Civita connection as well as terms with a factor $\partial_a (1/\sqrt{N_g})$.

The Lagrangian (\ref{lo}) describes an unusual  quantum
 geometry in the guiding center space with a flat time. The emergent geometric field equation looks  highly nontrivial, which reads
 \begin{eqnarray}
R+\frac{1}{2}\epsilon^{ab}\epsilon^{cd}\partial_a\partial_c g_{bd} + \frac{\sqrt{2}}{J(e)\sqrt{\theta g_{012}}}\partial^a({\cal G}_{aa}{\cal C})=0.~~ \label{gfe}
\end{eqnarray}
In the Chern-Simons gauge field parametrization, Eq. (\ref{gfe}) is simplified to the constraint equation
\begin{eqnarray}
\epsilon^{ab}\partial_aa_b-\theta\epsilon^{ab}\epsilon^{cd}\partial_aa_c\partial_b a_d/2=0, \label{ceq}
\end{eqnarray}
which is the first-order expansion in $\theta$ of Eq. (\ref{209}). Solving this equation provides a solution of the geometric field equation (\ref{gfe}).

\subsection{ Guiding center spin }
Now we discuss the effect of the Chern-Simons field in $J(e)$. Define the operator $\bar{\bf s}$ by
\begin{eqnarray}
\bar{\bf s}=\theta^2a_1a_2/g_{012}=-\delta_c x^1\delta_c x^2/g_{012}, \label{sg}
 \end{eqnarray}
where  $\delta_c x^a=\theta \epsilon^{ab}a_b$.
 We call $\bar{\bf s}$ the {\it guiding center spin operator}  by the following argument. Recall the guiding center rotation generator $L_{rot}$ defined by $L_{rot}=g_{ab}\hat{\Lambda}^{ab}$ \cite{haldane},  where $\hat\Lambda^{ab}(x)=\frac{1}{4l^2_B}\sum_i\{(\hat{Y}_i^a-x),(\hat{Y}_i^b-x)\}$. The Lie algebra of the $\hat\Lambda^{ab}$ operators is $sl(2,R)$.  Similarly, in the continuum theory, we consider the following generators:
 \begin{eqnarray}
 \Lambda^{ab}&=&-\frac{1}2\int d^2y\rho_0{\delta_c x^a(y)\delta_c x^b(y)},  \label{gc}
 \end{eqnarray}
{We can check that, $\Lambda^{ab}$ is symmetric, i.e., $\Lambda^{ab}=\Lambda^{ba}$, and after the lowest Landau level projection, we shall consider the Dirac bracket between the $\Lambda$s instead of the ordinary commutation relation. We can calculate their Dirac brackets explicitly}, for example,
\begin{eqnarray*}
	\{\Lambda^{11},\Lambda^{22}\}_{D}=\frac{i}{4}\int d^2y \frac{\theta^3}{2\pi}\epsilon^{ab}\partial_a(a_2a_2)\partial_b(a_1a_1) .
\end{eqnarray*}	
Now we use the constraint equation (\ref{ceq}) and the fact that the density of elections $\rho_0=\epsilon^{ab}\partial_a a_b$, and we get
    \begin{eqnarray}
    \{{\Lambda}^{11},{\Lambda}^{22}\}_{D} &=&\frac{i}{2}\int d^2y \frac{\theta^2}{4\pi}(4a_1a_2\theta\epsilon^{ab}\partial_aa_2\partial_ba_1)\nonumber\\
    &=&-\frac{i}{2}({\Lambda}^{12}\epsilon^{12}+{\Lambda}^{12}\epsilon^{12}+{\Lambda}^{12}\epsilon^{12}+{\Lambda}^{12}\epsilon^{12}).\nonumber\\
    \end{eqnarray}
Similarly the results for $\{{\Lambda}^{11},{\Lambda}^{12}\}_{D}$ and $\{{\Lambda}^{22},{\Lambda}^{12}\}_{D}$ can be written in a compact form:
\begin{eqnarray}
\{\Lambda^{ab},\Lambda^{cd}\}_{D}=-\frac{i}2({\Lambda}^{ac}\epsilon^{bd}+  {\Lambda}^{bd}\epsilon^{ac}+(a\leftrightarrow b)). \label{sl2}
\end{eqnarray}
 Equation (\ref{sl2}) is the Lie algebra of $sl(2,R)$, same as given in \cite{haldane}. Therefore, we can interpret the ${\Lambda}^{ab}$ operators as the continuum analog to Haldane's $\Lambda^{ab}$. 
 The guiding center spin $\bar s$ was defined by the expectation value of $\Lambda^{ab}$ in the Laughlin state \cite{haldane}:
 \begin{eqnarray}
 \lim_{N\to\infty}\langle \Lambda^{ab}\rangle/N =\bar s\, g_0^{ab}/2. \label{spinspin}
 \end{eqnarray}
 Therefore,  $\langle\bar {\bf s}\rangle=\bar s$ since $g_0^{12}=-g_{012}$. In our continuum formulation, we use $\bar s$ to approximate $\bar{\bf s}$ and consider $\langle a_b\rangle=0$. Thus, Eq. (\ref{f}) can be approximated
by
 \begin{eqnarray}
\langle J(e) \rangle/4\pi\nu\approx (L_m+\bar s)^2/2\pi L_m. \label{sgs}
 \end{eqnarray}
Then $\bar s$ indeed plays a role of `spin' added to the `orbital angular momentum' in the above way.  It is a topological invariant which characterizes the FQH state in addition to the filling factor  \cite{haldane,haldane2}. For $\nu=1/m$,
$L_m=m/2$, $\bar s=-(m-1)/2$ and $J_m=L_m+\bar s=1/2$ \cite{haldane}. Notice  that $g_{012}$ is zero for a rotationally invariant system. This is why the guiding center spin was not found in the previous studies based on the usual Laughlin wave function. However, the guiding center spin is still well-defined even for a system with rotational invariance because of the cancellation of $g_{012}$ in Eq. (\ref{sgs}). {A parallel analysis of the emergent quantum geometry and guiding center spin in the language of noncommutative geometry is presented in Appendix \ref{appen2} to show the consistency between the method of Dirac bracket and the noncommutative geometry methods. Through the area-preserving symmetry, the exotic quantum geometry is encoded in the guiding center coordinates $(y_1,y_2)$. One can reveal its noncommutative nature through the Moyal $*$-product or the Dirac bracket for the target-space field $x$ (or the {\it zweibein} fields $e$).}		}

 \section{ Possible generalization to other FQH states }
In the K-matrix theory of Abelian FQH  states, the electron is divided or decomposed into a set of particles with charge vector $t_A$ ($A=1,\cdots, K$) for $K$ being the dimensions of the K matrix \cite{effth2}.  The charge vector relates to the filling factor via
$\nu=t^AK^{-1}_{AB}t^B$. The particle positions are $x_{aA}$. Then a Hubbard-Stratonovich-like transformation can be applied to the single electron Lagrangian (\ref{L0}):
\begin{equation}
L_{e,0}=\epsilon^{ab}(\dot{x}^A_a-\dot{x}_aT^A)K_{AB}(x^{B}_b-x_bT^B)-\epsilon^{ab}\dot{x}_a x_b/2.\nonumber
\end{equation}
The vector $T_A$ is chosen so that $T^AK_{AB}T^B=1/2$; i.e., it relates to the charge vector by  $T_A=K^{-1}_{AB}t^B/\sqrt{2\nu}$.
 $L_{e,0}$ then reads
\begin{equation}
L_{e,0}=\epsilon^{ab}\dot{x}_a^AK_{AB}x^B_b-\epsilon^{ab}\dot{x}_a^AK_{AB}T^Bx_b-\epsilon^{ab}\dot{x}_aT^AK_{AB}x^B_b .
\nonumber
\end{equation}
Using the equation of motion of the $x^A_a$ and the symmetric property of the K matrix, we have
\begin{equation}
\dot{x}^A_a-\dot{x}_aT^A=0\label{16}.
\end{equation}
Putting this solution into $L_{e,0}$ and dropping out a total time
derivative term, we have
\begin{equation}
L_{e,0}=\epsilon^{ab}\dot{x}^A_aK_{AB}x^B_b.\label{17}
\end{equation}
In this way, we transform a single electron problem to a free theory of particles described by K-matrix Chern-Simons mechanics. If we also have a variational metric $g_0$ for these K-matrix FQH states using pseudopotentials, we can follow the similar track that we have used to obtained the Laughlin state to obtain a K-matrix Chern-Simons gauge theory. Finally we will have the Lagrangian of the K-matrix Chern-Simons gauge theory:
 \begin{eqnarray}
L_{K,0}=\frac{K_{AB}}{4\pi }\int d^2y(\epsilon^{\mu\nu\rho} a^A_\mu\partial_\nu a_\rho^B+\theta\epsilon^{ab}\epsilon^{ij}a_0\partial_a a^A_i\partial_b a^B_j). \nonumber
\end{eqnarray}
Similarly, we can discuss the shift and guiding center spin of these states as before. From the Lagrangian (\ref{loother}), we have
\begin{eqnarray}
L_0&=&\int_{S^2}d^2y (\frac{K_{AB}}{4\pi}\epsilon^{\mu\nu\rho}a_\mu^A\partial_\nu a_\rho^B+\frac{1}{2\pi}s_C\epsilon^{ab}\Omega_a\partial_ba_{0}^C\nonumber\\
&+&\frac{\theta^2K_{AB}a_1^Aa_2^B}{g_{012}}\epsilon^{ab}\Omega_a\partial_ba_{0}+\cdot\cdot\cdot).
\end{eqnarray}
Here we introduce the spin vector $s_C$ as defined in \cite{wenzee}, the shift $S$ will be $S=2(tK^{-1}s_c)/\nu$ on sphere. Now let us focus on the guiding center spin term $\langle \frac{\theta^2K_{AB}a_1^Aa_2^B}{g_{012}} \rangle$ as in (\ref{spinspin}). If we take the elementary droplet point of view, and borrow Haldane's expression for the guiding center spin\cite{haldane}, it is determined by
\begin{equation}
\bar{s}=\lim_{\bar{N}\rightarrow\infty}\frac{1}{\bar{N}}\sum_{m=0}^{q\bar{N}-1}L_m(n_m(\bar{g},\vec{r})-\nu), \quad L_m=(m+\frac{1}{2}),\label{sspin}
\end{equation}
where $n_m(\bar{g},\vec{r})$ is the occupation of the guiding center orbitals and $L_m$ is the corresponding angular momentum. For an "electron-type" $\nu=p/q$($p<q$) state, the elementary droplet has $q$ orbitals with the first $p={\rm rank}~K$ orbitals filled, and the total angular momentum will be,
\begin{equation}
L_{tot}^{e}=\sum_{n=0}^{p-1}\frac{2n+1}{2}=\frac{p^2}{2}.
\end{equation}
The reference angular momentum is given by assigning each $q$ orbital with a factor $p/q$,
\begin{equation}
L_{ref}^{e}=\frac{p}{q}\sum_{n=0}^{q-1} \frac{2n+1}{2}=\frac{pq}{2}.\label{refang}
\end{equation}
Therefore the guiding center spin $\bar{s}$ for the electron-type state is,
\begin{equation}
\bar{s}^{e}=L_{tot}^{e}-L_{ref}^{e}=\frac{p}{2}(p-q)=\frac{pq}{2}(\nu-1).
\end{equation}
For a  "hole-type" $p/q$ state, the elementary droplet also has $q$ orbitals but with the last $p={\rm rank}~K $ orbitals filled, the total angular momentum will be
\begin{equation}
L_{tot}^{h}=\sum_{n=q-p}^{q-1}\frac{2n+1}{2}=\frac{2pq-p^2}{2}.
\end{equation}
The reference angular momentum is the same as in Eq. (\ref{refang}):
\begin{equation}
L_{ref}^{h}=\frac{p}{q}\sum_{n=0}^{q-1} \frac{2n+1}{2}=\frac{pq}{2}.
\end{equation}
Then the guiding center spin $\bar{s}$ for the hole-type state is,
\begin{equation}
\bar{s}^{h}=L_{tot}^{h}-L_{ref}^{h}=\frac{p}{2}(q-p)=\frac{pq}{2}(1-\nu).
\end{equation}
These two results for guiding center spin are also in coincidence with (\ref{sspin}) if we take the configuration of the elementary droplet as before. From these two results, we notice the following properties of guiding center spin\cite{haldane2}:

(1) it is odd under particle-hole transformation,

(2) it is negative for electron-type states and positive for hole-type states,

(3) it is zero for empty and full filled Landau levels. \\
To make these statements clearer, we present some examples. For the hole-type $\nu=2/3$ state,  ${\rm rank}~K=2$, there are three orbitals in the elementary droplet and the last two orbitals are filled, or the occupation function $n_m$ is $n_0=0$ and $n_1=n_2=1$ as defined in Eq.(\ref{sspin}). The corresponding guiding center spin is
\begin{equation}
\bar{s}=L_{tot}-L_{ref}=(\frac{3}{2}+\frac{5}{2})-\frac{2}{3}(\frac{1}{2}+\frac{3}{2}+\frac{5}{2})=1.
\end{equation}
Similarly, for the electron-type $\nu=2/5$ state, ${\rm rank}~K=2$, the elementary droplet consists of five orbitals with the first two orbitals filled. The occupation function $n_m$ is $n_0=n_1=1$ and $n_2=n_3=n_4=0$, and the guiding center spin is
\begin{equation}
\bar{s}=L_{tot}-L_{ref}=(\frac{1}{2}+\frac{3}{2})-\frac{2}{5}(\frac{1}{2}+\frac{3}{2}+\frac{5}{2}+\frac{7}{2}+\frac{9}{2})=-3.
\end{equation}
The above examples are the same as those given by Haldane\cite{haldane2}.

\vspace{5pt}
\section{ Conclusions }
We identify the electron position fluctuation around its guiding center in a given Laughlin state with the collective dynamic internal geometric fluctuation {which is the origin of the gauge fields in the Chern-Simons theory}. {By using the Dirac bracket method,} we {show} that the noncommutative Chern-Simons theory is a better description {than the usual commutative Abelian Chern-Simons theory} in the lowest Landau level. There is a quantum geometry emerging from the Chern-Simons gauge fluctuations. The shift and guiding center spin were naturally defined. We have used the zero mass limit to do the lowest Landau level projection. Therefore, the application to higher Landau level physics remains open. We discuss the possible generalization to other fractional quantum Hall states with the emergence of K-matrix formalism and its guiding center spin. The even denominator filling factor FQH states are beyond our reach at this moment.

\acknowledgments
The authors thank Zheng-Cheng Gu,  Long Liang and Kun Yang for useful discussions.  X. L. is grateful to the warm hospitality of the Department of Physics of Fudan University where this work was completed. This work was supported in part by the 973 program of the MOST of China (Grant No. 2012CB821402), the NNSF of China (Grants No. 11174298 and No. 11474061), and the United States National Science Foundation through Grant No. PHY-1068558.

\appendix

\section{A brief introduction to noncommutative geometry}\label{appen1}

{In this appendix, we demonstrate that by starting with a noncommutative geometric approach to the guiding center space, which uses merely the usual star product, one can achieve the 
same (noncommutative) Chern-Simons theory without using Dirac brackets. This implies that the Dirac bracket method and that of noncommutative geometry are parallel and consistent with each other. \\
\\
In order to make our paper self-contained, first we shall give a brief introduction to non-commutative geometry\cite{liwu,wunc}.} On a two-dimensional noncommutative space, the coordinates satisfy a Heisenberg-like commutation relation,
\begin{equation}
[x_a,x_b]=i\theta\epsilon_{ab}, \label{203}
\end{equation}
where $\theta$ is a constant and $\epsilon_{ab}$ is the Levi-Civita symbol. We can interpret this commutation relation in two ways\cite{wunc}. One way is to think $x_a$ as operators in a Hilbert space and they satisfy the noncommutative relation (\ref{203}). From this point of view, the noncommutative space can be interpreted as a generalization of the { phase space} in the usual quantum mechanics. The other way is the deformation quantization\cite{hweyl,hjg}. We can regard the coordinates $x_a$ as ordinary functions and they generate a noncommutative algebra of functions on the space. In other words, we are able to develop a classical field theory from this perspective. The fields themselves are ordinary functions of formally commuting variables $\{x_a\}$, but the local products in the field algebra are defined by (or deformed to) the Moyal $*$-product\cite{jemo},
\begin{equation}
f(x)*g(x)=\exp\left[\frac{i}{2}\theta\epsilon^{ab}\frac{\partial}{\partial \xi^a}\frac{\partial}{\partial \eta^b}\right]f(x+\xi)g(x+\eta)|_{\xi=\eta=0}.\label{204}
\end{equation}
Therefore, the commutator is defined by using the $*$-product:  
\begin{equation}
[f,g]_{*}\equiv f*g-g*f.
\end{equation}
Using these definitions, one can easily check,
\begin{equation} 
[x_a,x_b]_{*}=x_a*x_b-x_b*x_a=i\theta\epsilon_{ab}. \label{205}
\end{equation}

With the help of the $*$-product, the action principle and equations of motion can be straightforwardly generalized to noncommutative geometry\cite{liwu}. The key difference from the usual field theory is that whenever we have the multiplication between two fields, we shall use the $*$-product (\ref{204}) instead of the usual product. For instance, we shall write the scalar $\phi^4$ interaction as $\phi*\phi*\phi*\phi$. Although we usually will have more interaction vertices in a noncommutative field theory\cite{wunc}, the higher-order (derivative) terms are controlled by the $*$-product and organized in a mathematically neat way. The noncommutativity of the spacial coordinates (\ref{205}) implies that there is a minimal uncertainty area,
\begin{equation}
\Delta x^1\Delta x^2\sim \theta. 
\label{a206}
\end{equation}
The minimal uncertainty area has a physical meaning in fractional quantum Hall effect; namely, it is the area
occupied by a single electron\cite{nccs}. This relation actually give us some kind of regularization, like a lattice constant in a lattice field theory. Therefore when we calculate the quantum amplitudes, we choose a regularization that is consistent with the one imposed by the minimal uncertainty area. Another profound feature of this minimal uncertainty area or of any noncommutative field theory, is the UV-IR entanglement\cite{wunc,mrs,chenwu}.

\section{ Guiding center spin }\label{appen2}
{If we expand the noncommutative Lagrangian (\ref{al00}) to the first order in the noncommutative parameter $\theta$, we will have the same results as in the Dirac bracket approach, namely, the Lagrangian (\ref{l00}). Therefore the discussions of the emergent geometry will be similar to those in Sec. III which we will not repeat here. The only difference we shall emphasize is in the $\Lambda^{ab}$ operators which is needed to define the guiding center spin in the language of noncommutative geometry. As before,} the guiding center operator $\bar{\bf s}$ is defined by
\begin{eqnarray}
\bar{\bf s}=\theta^2a_1a_2/g_{012}=-\delta_c x^1\delta_c x^2/g_{012}, \label{asg}
\end{eqnarray}
where  $\delta_c x^a=\theta \epsilon^{ab}a_b$.
{In Ref. \cite{haldane}, Haldane defined the guiding center rotation generator $L_{rot}$ by $L_{rot}=g_{ab}\hat{\Lambda}^{ab}$,  with $\hat\Lambda^{ab}(x)=\frac{1}{4l^2_B}\sum_i\{(\hat{Y}_i^a-x),(\hat{Y}_i^b-x)\}$. The  $\hat\Lambda^{ab}$ operators form an $sl(2,R)$ algebra.  Similarly, in the description of noncommutative geometry, we consider the following generators:}
\begin{eqnarray}
\Lambda^{ab}&=&-\frac{1}2\int d^2y\rho_0{\delta_c x^a(y)*\delta_c x^b(y)},  \label{agc}
\end{eqnarray}
Let us write down all the components of $\Lambda^{ab}$ to the first order of $\theta$,
\begin{eqnarray}
\Lambda^{11}&=&-\frac{\theta^2}{2}\int d^2y a^2a^2,\\
\Lambda^{12}&=&\frac{\theta^2}{2} \int d^2y (a^1a^2+\frac{1}{2}\theta \epsilon^{ab}\partial_a a_2 \partial_b a_1),\\
\Lambda^{22}&=&-\frac{\theta^2}{2}\int d^2y a^2a^2,\\
\Lambda^{21}&=&\frac{\theta^2}{2} \int d^2y (a^1a^2-\frac{1}{2}\theta \epsilon^{ab}\partial_a a_2 \partial_b a_1).
\end{eqnarray}
We shall also notice that, because of the constraint equation (\ref{ceq}), we have,
\begin{equation}
\int d^2y \theta\epsilon^{ab}\partial_a a_2 \partial_b a_1 = \int d^2y \epsilon^{ab}\partial_aa_b= \int d^2y \rho_0=N,
\end{equation}
where $N$ is the total particle number of electrons. Therefore
\begin{eqnarray}
\Lambda^{12}&=&\frac{\theta^2}{2} (\int d^2y a^1a^2+\frac{N}{2}),\\
\Lambda^{12}&=&\frac{\theta^2}{2} (\int d^2y a^1a^2-\frac{N}{2}).
\end{eqnarray}
We can symmetrize $\Lambda^{12}$ and $\Lambda^{21}$ by defining $\tilde{\Lambda}^{12}=\tilde{\Lambda}^{21}\equiv\frac{1}{2}(\Lambda^{12}+\Lambda^{21})$. We can also calculate the noncommutative brackets between $\tilde{\Lambda}^{ab}$ ($\tilde{\Lambda}^{11}\equiv\Lambda^{11}$, $\tilde{\Lambda}^{22}\equiv\Lambda^{22}$).
Because $\tilde{\Lambda}^{ab}=\tilde{\Lambda}^{ba}$ and $[\tilde{\Lambda}^{ab},\tilde{\Lambda}^{cd}]=-[\tilde{\Lambda}^{cd},\tilde{\Lambda}^{ab}]$, there are only three independent commutators, i.e., $[\tilde{\Lambda}^{11},\tilde{\Lambda}^{22}]$, $[\tilde{\Lambda}^{11},\tilde{\Lambda}^{12}]$, and $[\tilde{\Lambda}^{22},\tilde{\Lambda}^{12}]$. We can calculate all of them explicitly. For example,
\begin{eqnarray*}
	[\tilde{\Lambda}^{11},\tilde{\Lambda}^{22}]_{*}=\frac{i}{4}\int d^2y \frac{\theta^3}{2\pi}\epsilon^{ab}\partial_a(a_2a_2)\partial_b(a_1a_1) .
\end{eqnarray*}
Using the constraint equation (\ref{ceq}), we notice that the density of elections $\rho_0=\epsilon^{ab}\partial_a a_b$, and we get
\begin{eqnarray}
[\tilde{\Lambda}^{11},\tilde{\Lambda}^{22}]_{*} &=&\frac{i}{2}\int d^2y \frac{\theta^2}{4\pi}(4a_1a_2\theta\epsilon^{ab}\partial_aa_2\partial_ba_1)\nonumber\\
&=&-\frac{i}{2}(\tilde{\Lambda}^{12}\epsilon^{12}+\tilde{\Lambda}^{12}\epsilon^{12}+\tilde{\Lambda}^{12}\epsilon^{12}+\tilde{\Lambda}^{12}\epsilon^{12}).\nonumber\\
\end{eqnarray}
Similarly we can calculate $[\tilde{\Lambda}^{11},\tilde{\Lambda}^{12}]_{*}$ and $[\tilde{\Lambda}^{22},\tilde{\Lambda}^{12}]_{*}$, and we have,
\begin{eqnarray}
[\tilde{\Lambda}^{ab},\tilde{\Lambda}^{cd}]_{*}=-\frac{i}2(\tilde{\Lambda}^{ac}\epsilon^{bd}+  \tilde{\Lambda}^{bd}\epsilon^{ac}+(a\leftrightarrow b)). \label{asl2}
\end{eqnarray}
Equation (\ref{asl2}) is the Lie algebra of $sl(2,R)$, the same as given in \cite{haldane} {and also consistent with Eq. (\ref{sl2})}. Therefore we can also interpret the $\tilde{\Lambda}^{ab}$ operators as the continuum analog to Haldane's $\Lambda^{ab}$ {and have the same interpretation of the guiding center spin as in Eq. (\ref{spinspin})}. If we use the original asymmetric $\Lambda^{ab}$, we still have the $sl(2,R)$ algebra for the noncommutative bracket $[\Lambda^{ab},\Lambda^{cd}]_{*}$. 


\end{document}